\documentclass{pasj00}

\usepackage{natbib}

\newcommand{\kms}{km~s$^{-1}$}
\newcommand{\ion}[2]{#1\,{\sc #2}}
\newcommand{\lam}{$\lambda$}
\newcommand{\ecss}{erg~cm$^{-2}$~s$^{-1}$~sr$^{-1}$} 
\newcommand{\hinode}{\emph{Hinode}}
\newcommand{\as}{${^\prime}{^\prime}$}

\newcommand{\axii}{\emph{A193}}
\newcommand{\aix}{\emph{A171}}
\newcommand{\axviii}{\emph{A94}}
\newcommand{\axxi}{\emph{A131}}
\newcommand{\axiv}{\emph{A211}}
\newcommand{\axvi}{\emph{A335}}
\newcommand{\ahe}{\emph{A304}}

\begin{document}
\SetRunningHead{P.R.~Young \& K.~Muglach}{A Coronal Hole Jet Observed
  with Hinode and SDO}

\title{A Coronal Hole Jet Observed with Hinode and the Solar Dynamics
  Observatory} 

\author{Peter R. \textsc{Young}}
\affil{College of Science, George Mason University, 4400 Unversity
  Drive, Fairfax, VA 22030, USA}
\email{pyoung9@gmu.edu}
\and
\author{Karin  \textsc{Muglach}}
\affil{Code 674, NASA Goddard Space Flight Center,
    Greenbelt, MD 20771, USA}
\affil{ARTEP, Inc., Ellicott City, MD 21042, USA}
\email{kmuglach@gmx.de}


\maketitle

\begin{abstract}
A small blowout jet was observed at the boundary of the south polar
coronal hole on 2011 February 8 at around 21:00~UT. Images from the
Atmospheric Imaging Assembly (AIA) on board the Solar Dynamics
Observatory (SDO) revealed an expanding loop rising from one footpoint
of a compact, bipolar bright point. Magnetograms from the Helioseismic
Magnetic Imager (HMI) on board SDO showed that the jet was triggered
by the cancelation of a parasitic positive polarity feature near the
negative pole of the bright point. The jet emission was present for
25~mins and it extended 30~Mm from the bright point. Spectra from the
EUV Imaging Spectrometer on board \hinode\ yielded a temperature and density of
1.6~MK and 0.9--1.7 $\times$ $10^8$~cm$^{-3}$ for the ejected
plasma. Line-of-sight velocities reached up to 250~\kms\ and were
found to increase with height, suggesting plasma acceleration within
the body of the jet. Evidence was found for twisting motions within
the jet based on variations of the LOS velocities across the jet
width. The derived angular speed was in the range 9--$12\times
10^{-3}$~rad~s$^{-1}$, consistent with previous measurements from jets.
The
density of the bright point was 7.6 $\times$ $10^8$~cm$^{-3}$, and the
peak of the bright point's emission measure occurred at 1.3~MK, with
no plasma above 3~MK.
\end{abstract}

\section{Introduction}

Coronal jets produce hot plasma, $>$~1~MK, that is ejected upwards
away from the solar surface at high speeds up to several hundred
\kms, and they have been identified in coronal holes, quiet Sun and
active regions \citep{shimojo96}. The distinctive shape of jets is
believed to be due to small-scale closed magnetic loops interacting
with larger-scale open field structures \citep{shibata86} and, as
such, 
they are important as a means for studying basic plasma
heating and acceleration mechanisms. All jets have their origins in a
bright point on the solar disk. Often the bright point will have a
simple, bipolar morphology (particularly in the case of coronal hole
jets) and the coronal loops of the bright point are referred to as the
base arch. The name `jet' is sometimes used to refer to the entirety of
the event, with the term `spire' used to refer to the plasma ejected
upwards from the bright point.

Recently \citet{moore10} suggested the existence of two types of
coronal hole jet, referred to as \emph{standard} and \emph{blowout}. The distinction
was further clarified in \citet{moore13}, who identified standard jets
as having a single, narrow spire and a brightening at the edge of the
bright point. Blowout jets have more complex spires (often a `curtain'
of emission is seen) and the interior of the bright point
intensifies. Subsequently, a number of authors have attempted to
interpret their jet observations (both inside and outside of coronal
holes) in terms of this dichotomy and it is
apparent that a number of jets show complex behavior not easily
assigned to either group, particularly when the high spatial and
temporal resolution data from the Atmospheric Imaging Assembly (AIA)
on board the Solar Dynamics Observatory (SDO) are used. The blowout
jets have been interpreted as being small-scale coronal mass ejections
\citep[micro-, or mini-CMEs,][]{nistico09,hong11}, with cool plasma from a mini-filament ejected
with the coronal plasma, although this was not a defining
characteristic for \citet{moore13}, who classified events according to
features seen in X-ray observations.

The launch of SDO in 2010 has
been a boon for studies of transient events such as jets, as it yields
full-disk solar images in multiple wavelengths at regular time
cadences and with high spatial 
resolution. This ensures that any event on the Sun is almost certain
to be captured with a consistent, high quality data-set. Jets seen
with AIA often show complex ejecta \citep{liu11, shen11, chen12,
  shen12, kslee13, schmieder13} and categorization into standard or
blowout may not be straightforward \citep{liu11,schmieder13}.  An
important measurement that has been made from some SDO events is a
twisting motion of the spire, identified as motions of compact bright
features within the ejecta that can be tracked to reveal motions transverse
to the jet axis \citep{liu11,shen11,chen12,hong13}. Twist is an
important parameter for understanding the physical mechanism behind
the jet evolution as it can arise from emerging flux \citep{pariat09},
and \citet{hong13} have suggested that twist in blowout jets could
arise from the erupting mini-filament associated with the event.

Most of the jets studied with SDO data have been
large events seen in the quiet Sun or near active
regions. Coronal holes hold a particular advantage for studying jets as the coronal
emission above 1~MK is weak, and so ejecta that are hotter than this
can be observed clearly. Additionally the magnetic structure in the
photosphere is generally simpler, with the magnetic poles of the
bright point lying within the unipolar magnetic field of the coronal hole.
\citet{shen11} and \citet{hong13} reported jets seen in polar
coronal holes, but in each case the jet was at the limb and so
magnetic field data were unavailable.
\citet{young-blowout} presented SDO observations of a blowout jet occurring in
a coronal hole on the disk, combined with EUV spectroscopic
measurements from the EUV Imaging Spectrometer on board \hinode.
The jet was triggered by the two
dominant polarities of the bright point converging and canceling with
each other. The jet's spire was a broad curtain that extended over
70~Mm, and intense, small-scale kernels were seen within the bright
point as the jet evolved. The spectroscopic data enabled the
temperature and density of the jet plasma to be estimated at 1.4~MK and
$2.7\times 10^8$~cm$^{-3}$, respectively. The LOS jet speed was
shown to reach up to 250~\kms\ and was found to increase with height.

The present work is a follow-up to that of \citet{young-blowout}, using
data from the same observation set but for a different event occurring
at the coronal hole boundary. This event showed similarly large LOS
velocities, but it was much smaller, it showed a different magnetic
field evolution, and the morphology of the ejected plasma was
different. Parameters of the jet measured from the AIA and
Helioseismic Magnetic Imager (HMI) onboard SDO  and \hinode/EIS are
presented in Sects.~\ref{sect.sdo} and \ref{sect.eis}. A summary is
given in Sect.\ref{sect.summary} with a particular focus on
comparisons with the jet of \citet{young-blowout}.

\section{Observations}\label{sect.obs}

Hinode Observing Program No. 177 (HOP 177) was run during 2011
February 8--10, giving a continuous observation of the south coronal
hole. The EIS instrument obtained large format rasters covering an
area of 179\as\ $\times$ 512\as\ at a 62~minute cadence, and
Dopplergrams formed from the \ion{Fe}{xii} \lam195.12 emission line
(formed at 1.5~MK)
revealed 35 large-scale, blue-shifted structures. An event captured on
February 9 
was classed as a blowout jet and was presented in
\citet{young-blowout}. 
It was one of only two events that showed blue-shifted velocity
components at speeds $> 150$~\kms, and the second is described in the
current work. As the latter occurred on February 8, we refer to it as
the ``8-Feb'' jet, and the 
former as the ``9-Feb'' jet.

The low time cadence of the EIS rasters meant that the jet was only
captured in a single raster, so time evolution can not be
studied. However, the AIA and HMI instruments on board SDO yielded
coronal images and photospheric magnetic field measurements at a high
cadence. AIA has seven EUV filters and images are obtained at a
12-second cadence. In this paper we use the shorthand \axii\ to
indicate the AIA filter centered at 193~\AA. The AIA filters can have
a complex response to the solar plasma temperature, depending on solar
conditions and the emission lines that contribute to the
bandpasses \citep{odwyer10}. For the present coronal hole observation,
the \axxi\ channel is dominated by \ion{Fe}{viii} emission (0.5~MK),
the \aix\ channel is dominated by \ion{Fe}{ix} emission (0.8~MK),
\axii\ by \ion{Fe}{xi} and \ion{Fe}{xii} emission (1.4--1.6~MK), and \axiv\
by \ion{Fe}{xi} emission (1.4~MK).

LOS magnetograms are taken from the HMI instrument, for which the data
product pipeline yields 12-minute and 45-second cadence magnetograms
(the former have a higher signal-to-noise). Further details of the
HOP~177 data-set are given
in \citet{young-blowout}.

\section{SDO observations}\label{sect.sdo}

\begin{figure*}
 \begin{center}
  \includegraphics[width=15cm]{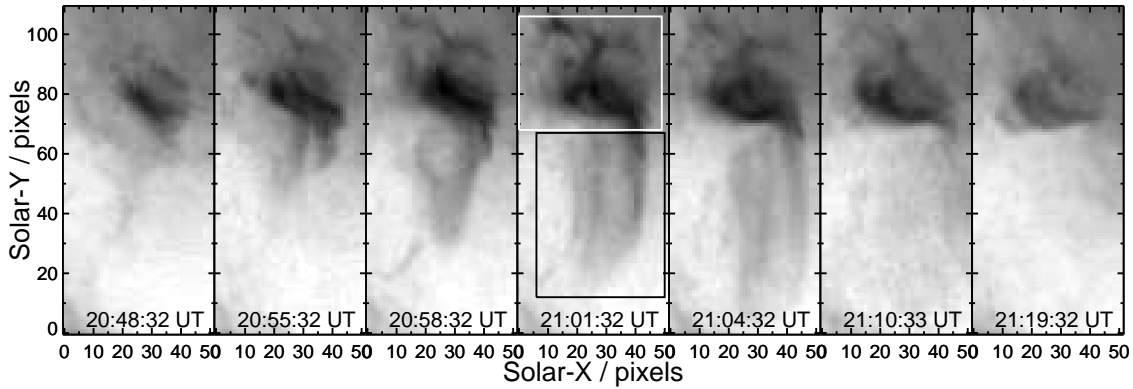}
 \end{center}
\caption{A set of image frames from the \axii\ filter. The logarithm
  of the intensity is shown, with a reversed color table. One pixel
  corresponds to 0.6\as. The white and black boxes on the middle panel
indicate the spatial regions averaged to yield the light curves shown
in Figure~\ref{fig.lc2}.}\label{fig.a193}
\end{figure*}

The jet evolved  over the period 20:50 to 21:15~UT on 2011 February
8, erupting from a bright point at the south coronal hole boundary at
a latitude of $33^\circ$S. Movie~1 
shows the evolution of the jet in
five different AIA filters,
and Figure~\ref{fig.a193} shows 
seven \axii\ images of the event that show the bright point and the
ejected plasma. The images in the movie and Figure~\ref{fig.a193} were
obtained by averaging five consecutive 12-second cadence images. The
movie shows \ahe, \axxi, \aix, \axii\ and 
\axiv\ images at a 1-minute cadence from 20:40 to 21:19~UT. 
The
\axviii\ filter is not shown as there was  no emission from the jet, and the \axvi\ filter
showed only very weak emission from the bright point (at the level of
2--3~DN).

According to the classification of \citet{moore13}, the broad,
loop-like structure of the ejected plasma shown in
Figure~\ref{fig.a193} identifies the jet as the
blowout type. In addition, the whole of the bright point brightens
during the event, rather than just a small brightening at the side of
the bright point.

The loop-like shape of the jet (third and fourth panels of
Figure~\ref{fig.a193}) began rising at 20:50~UT, and it came from the
left-side of the bright point. \emph{It was not one of the base arch
  loops of the bright point.} This can best be seen in Movie~1, particularly by studying
the \aix\ and \axii\ images. In the second frame of
Figure~\ref{fig.a193}, the vertical structure coming from the middle
of the bright point is the \emph{right} leg of the loop; the vertical
structure at the right side of the bright point is distinct from the
loop.

In the initial rise phase of the loop, it emitted in \aix\ and \axii,
but from 20:57~UT it was seen in absorption in \aix. This can be
understood if the loop contains both cold chromospheric plasma and hot
coronal plasma. The \aix\ channel is dominated by \ion{Fe}{ix}, which
has a peak abundance at $\log\,T=5.90$ \citep[atomic data
from][]{chianti71}. If the plasma in the loop is heated to a
temperature of 1.5~MK (the temperature of formation of \ion{Fe}{xii})
as the loop rises, then the \ion{Fe}{ix} emission will fall by a
factor $\ge 10$ due to the higher ionization or iron ions: the loop
essentially disappears in the \aix\ channel. However, the cold
chromospheric plasma in the loop will absorb the coronal hole
\ion{Fe}{ix} emission that is \emph{behind} the loop, thus explaining
why the loop is seen in absorption in the \aix\ channel. The cold
ejected plasma is not seen in \ahe\ however, except in the earliest
stages of hte loop's rise. We believe that this is because the weak
loop emission is lost against the background coronal hole \ahe\
emission. This highlights a problem with using \ahe\ emission to
distinguish blowout and standard jets: whereas \ahe\ emission is
relatively easy to identify for the limb jets studied by
\citet{moore13}, it can be difficult to identify such emission for
on-disk jets.


The rise of the jet loop occurred at the same time that filamentary
structure appeared around the bright point to the north and east
(frames 2 to 5 of Figure~\ref{fig.a193}). These may have been due to new
magnetic connections forming with nearby magnetic features, or
pre-existing connections becoming activated.

\begin{figure*}
 \begin{center}
  \includegraphics[width=15cm]{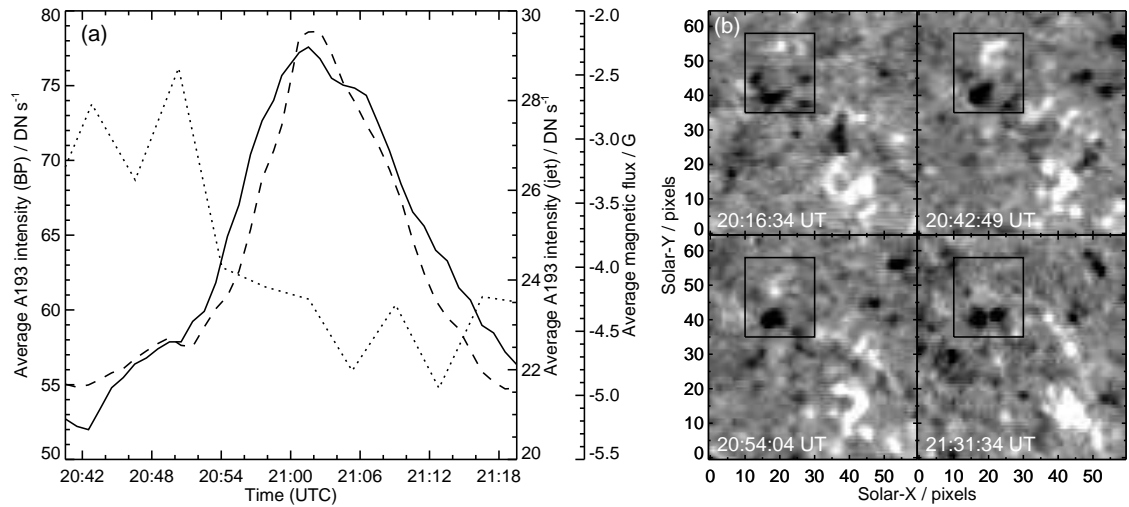}
 \end{center}
\caption{Panel a: the solid line shows the variation of the average \axii\
  intensity in the bright point (BP) with time. The dashed line shows
  the variation of the average \axii\ intensity of the jet, and the
  dotted line shows the variation of the average magnetic flux in the
  region indicated by the boxes in panel b. Panel b shows
  four HMI magnetograms for a region containing the bright point that
  yields the jet. The magnetograms have been saturated at levels of
  $\pm$30~G, and pixels have a size of 0.5\as.}\label{fig.lc2}
\end{figure*}

\axii\ light curves for the jet and bright point around the time of the jet
are shown in Figure~\ref{fig.lc2}. The bright point light curve was
obtained by averaging the intensity signal over an area of 30\as\
$\times$ 23\as, and the jet light curve was derived from a box of
27\as\ $\times$ 34\as\ (the boxes are indicated on the middle panel of
Figure~\ref{fig.a193}). The two
light curves are seen to rise and fall in unison, and the lifetime of
the jet is about 25~minutes. 

Comparing with the 9-Feb blowout jet there are clear differences in
the morphology and size of the ejected plasma. The 9-Feb jet had a
fan-shaped structure that extended for 70~Mm with a width of 30~Mm, whereas the 8-Feb
jet extended only 30~Mm and was 15~Mm wide. In addition there is no
evidence for the small, bright kernels within the bright point reported
by \citet{young-blowout}. There
were three small-scale brightenings that were present between 20:58 and
21:06~UT (the filamentary structure referred to earlier), and were most intense between 21:01 and 21:02~UT (fourth
frame of Figure~\ref{fig.a193}). However, these brightenings do not
occur within the base arch of the bright point; they do not become
brighter than the base arch; they do not show emission in the \axviii\
and 
\axvi\ filters; and they do not have such a sharply defined structure
as the 9-Feb jet kernels.

\begin{figure*}
 \begin{center}
  \includegraphics[width=12cm]{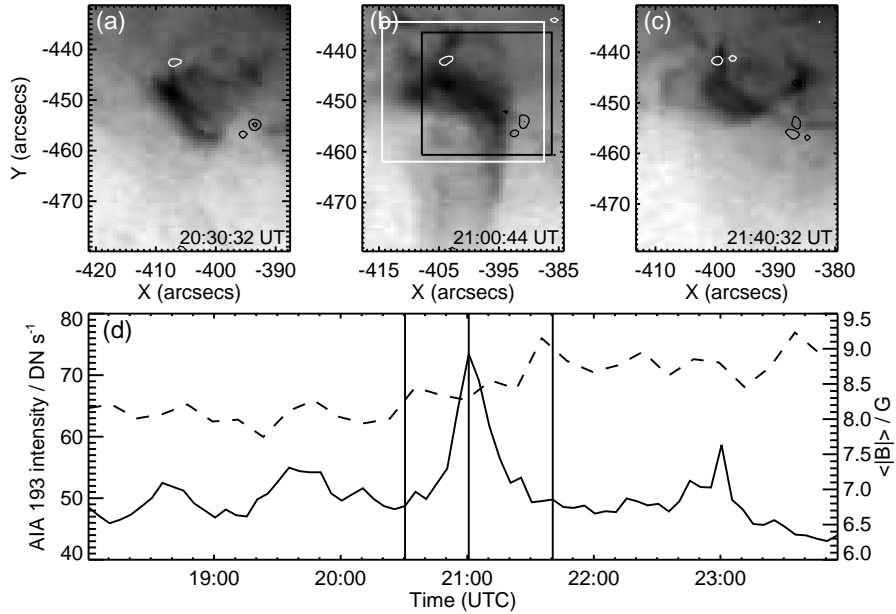}
 \end{center}
\caption{Panels a, b and c show \axii\ images with a reversed,
  logarithmic intensity scaling. Over-plotted are LOS magnetic field
  contours at levels of 75 and 150~G; white corresponds to negative
  polarity and black to positive polarity. The white and black boxes
  on panel b show the spatial areas that have been averaged to yield
  the \axii\ light curve and magnetic field variation shown in panel
  d. Panel d shows the \axii\
  light curve for the bright point (solid line), and  the variation of
the average unsigned magnetic flux for the bright point (dashed line). The vertical lines on
panel d indicate the times of the three images shown in panels a--c.}\label{fig.lc1}
\end{figure*}

With regard to the magnetic structure of the bright point, we first
consider the long term evolution as obtained from 12-minute cadence
HMI magnetograms and 5-minute cadence \axii\ images. (The \axii\
images were again derived by averaging five consecutive 12-second
cadence images.) Figure~\ref{fig.lc1}a--c shows three
\axii\ images from before, during and after the jet, with the HMI
magnetograms over-plotted as contours. These reveal that the bright
point had a fairly simple, tilted bipole structure that was largely unchanged
by the jet. 
In particular the strengths of the magnetic poles did not
change, nor did they move closer together. (The LOS magnetic field strengths of the negative and
positive poles were
around $-100$ and $130$~G at the time of the jet.) This contrasts with the
9-Feb jet for which the magnetic poles of the bright point came
together and canceled. For a six hour period from 18:00 to 24:00~UT,
 a \axii\ light curve for the bright point was created by averaging the
signal from a 28\as\ $\times$ 28\as\ box (indicated on Figure~\ref{fig.lc1}b), and it is shown in
Figure~\ref{fig.lc1}d. The basic morphology of the bright point first
appeared around 18:40~UT (the
first intensity peak in Figure~\ref{fig.lc1}d) and remained present
until 23:25~UT. There are four main intensity
brightenings seen in the light curve, the largest corresponding to the
jet discussed in the present work. The brightening at around 19:40~UT
also produced a jet, although much smaller and less dynamic than the
blowout jet. The evolution of the average, unsigned magnetic flux is
shown in Figure~\ref{fig.lc1}d as a dashed line. The flux was derived
by averaging the absolute flux over a box of size 22\as\ $\times$
25\as\ (indicated on Figure~\ref{fig.lc1}b) that enclosed the two dominant polarities of the bright
point. It can be seen that there was relatively little change in the
magnetic field, with a slight increase with time. This again
contrasts with the 9-Feb jet for which there was a clear decrease in
unsigned magnetic flux over a four hour period.

The fact that the loop ejected from the bright point 
first rose from the negative polarity part of the bright point gives
an important clue as to the magnetic evolution that triggered the
jet. To investigate further we took the 45-second cadence HMI
magnetograms and binned consecutive sets of five images to yield
magnetograms at a 225-second cadence. This sequence revealed that a positive
polarity feature with a strength of up to 35~G
appeared about 5\as\ north of the negative pole at around 20:10~UT, moved toward it, and
canceled with it by around 21:30~UT. Four frames from this sequence
are shown in  Figure~\ref{fig.lc2}b. 
The boxes on these plots show the location where flux cancelation
occurred, and the signed flux within these boxes was averaged to yield
the average magnetic flux that is shown in 
Figure~\ref{fig.lc2}a. As the weak positive polarity is canceled
around 20:54~UT, the average signed flux becomes more negative.
As discussed by \citet{young-blowout}, jets are commonly associated with
canceling flux. Unlike the 9-Feb jet, however, the dominant magnetic
poles of the bright point remained relatively unaffected, explaining why
the bright point was not destroyed by the jet in this case.  Note that
the high spatial resolution magnetograms of HMI are critical to
observing the weak positive polarity feature: a lower resolution or
lower sensitivity instrument likely would not have captured this.

To summarize the SDO observations, the jet was triggered by magnetic
cancellation at one footpoint of the bipole. The jet erupted as an
elongated loop-like shape visible only in the \axii\ and \axiv\
filters, suggesting that the plasma has a temperature of
1.5--2.0~MK. The ejection of chromospheric  plasma was implied by the loop
appearing in absorption in the \aix\ channel. The loop was present for about 25~mins before fading down
to background levels, and the light curve of the jet in the \axii\
filter was closely
matched to that of the bright point. The loop appeared highly elongated
with an aspect ratio up to approximately 4, although this may be a
line-of-sight effect. It is not clear if the loop broke open at the
apex during the eruption due
to the low signal-to-noise at large heights. The projected maximum extent of the
loop above the bright point was around 30~Mm.

\section{EIS observations}\label{sect.eis}

\begin{figure*}
 \begin{center}
  \includegraphics[width=15cm]{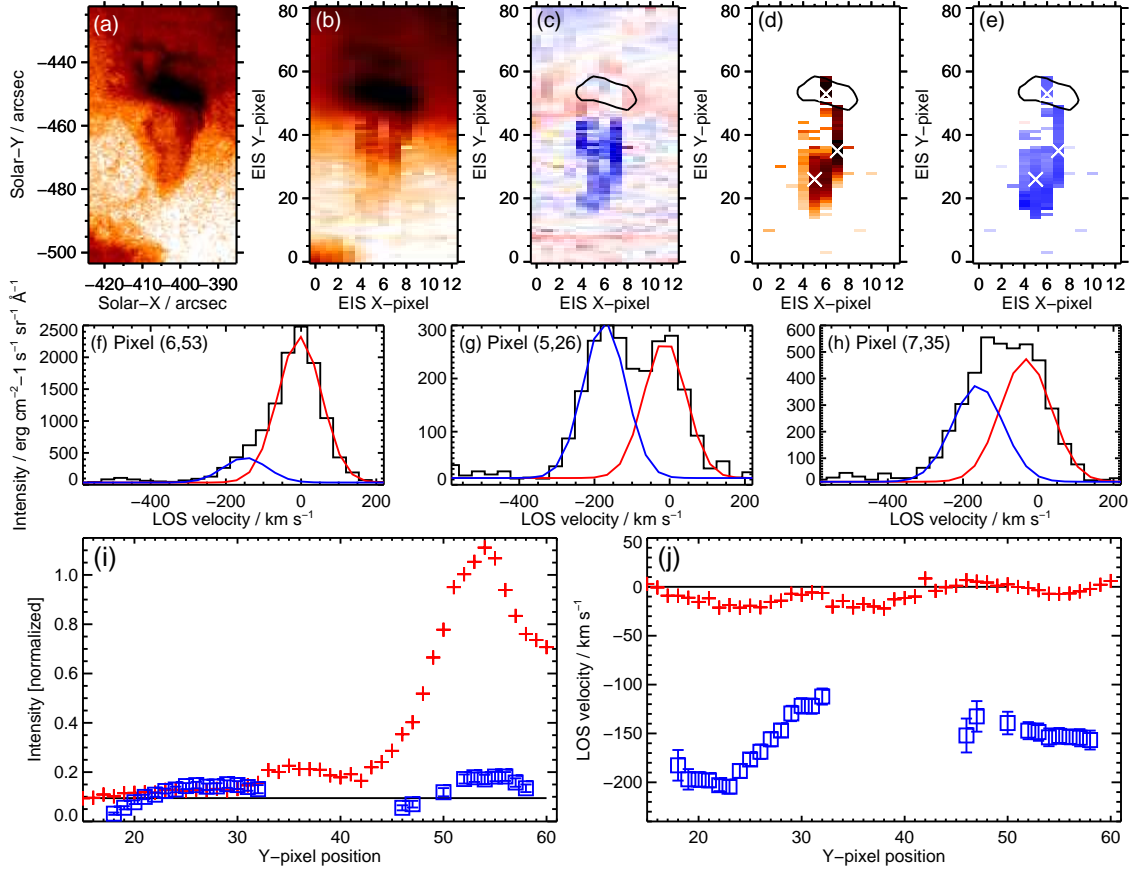} 
 \end{center}
\caption{Panel a shows an \axii\ image from 20:59~UT, with a
  reversed-log intensity scaling. Panels b and c show intensity and
  velocity images derived from the primary Gaussian component of 
  \ion{Fe}{xii} \lam195.12. A reversed-log intensity scaling is
  used for panel b, and panel c shows LOS velocities between $-20$ and
$+20$~\kms. Panels d and e show intensity and
  velocity images derived from the secondary Gaussian component of 
  \ion{Fe}{xii} \lam195.12. A reverse intensity scaling is used for
  panel d, and LOS velocities between $-200$ and $+200$~\kms\ are
  shown in panel e. For each of panels c--e, a black contour gives the
  location of the bright point. Panels f--h show line profiles from the spatial
  pixels identified with crosses on panels d and e. The black lines
  show the EIS spectrum, and the red and blue curves show the Gaussian
  fits for the primary (red) and secondary (blue) components. Panels i and j
  show intensity and velocity cross sections along X-pixel 6 for the
  primary (red) and secondary (blue) Gaussian components. The
  horizontal line on panel i shows the coronal hole background
  intensity level.}\label{fig.eis}
\end{figure*}

EIS
scanned the jet and bright point 
between 20:56 and 21:03~UT, and there was a strong velocity signal
in four consecutive exposures at 20:58, 20:59, 21:00 and 21:01~UT for which the
\lam195.12 line displayed significant emission on the short wavelength side of
the line. A
13 $\times$ 81 pixel sub-region around the jet  was extracted
and the \ion{Fe}{xii} \lam195.12 line was fit with two Gaussians over
this region using the IDL routine EIS\_AUTO\_FIT
\citep{eis_sw16}. The fit was constrained by requiring that the two
Gaussians had the same width, and that the spectrum background near
the lines was flat. 
We refer to the Gaussian component nearest the rest wavelength of the line as
the primary component, and the blue-shifted component as the secondary
component. 
Figures~\ref{fig.eis}b and d  show
intensity maps derived from the primary and secondary components. The
rest component looks quite similar to the A193 image 
(Figure~\ref{fig.eis}a), with a small bright point and weak, loop-like emission
extending southwards from the bright point.  The intensity image from
the secondary component (Figure~\ref{fig.eis}d)  has significant
intensity from the top of the weak loop, the right-hand side of the
loop and a section of the bright point (only the exposure at
20:59~UT, X-pixel 6). Velocity images from the two EIS \ion{Fe}{xii} components
are shown in Figures~\ref{fig.eis}c and e. The primary component of the line
shows significant blueshifts in the expanding loop, with
velocities of around $-10$ to $-30$~\kms. The secondary component
shows velocities of $-100$ to $-250$~\kms. 

Example line profiles from individual spatial pixels within the jet
are shown in Figures~\ref{fig.eis}f, g and h. Figure~\ref{fig.eis}f
shows a profile for the jet base where the secondary component is seen
as a bump on the side of the dominant primary component.
Figure~\ref{fig.eis}g shows an example profile from near the top of
the ejected loop where the secondary
component is stronger than the primary component,
and Figure~\ref{fig.eis}h shows an example where a single broad feature is
observed in the leg of the loop. Although two Gaussian components are fit to this feature, it
is possible that there is a wide range of velocities that combine to
give the observed feature.

Cuts through the EIS
intensity and velocity images are shown in 
Figures~\ref{fig.eis}i and j. X-pixel number 6 was selected,
corresponding to a time of
20:59~UT. This exposure shows a significant blue-shifted component in
the bright point (see also Figure~\ref{fig.eis}f) as well as the expanding loop. The horizontal line in
Figure~\ref{fig.eis}i shows the background coronal hole intensity (note: all
intensities have been normalized to the average intensity of the
bright point). The expanding loop intensity is only a factor two or
less brighter than the background, but because EIS is able to resolve
the two velocity components of \lam195.12 then a faint,
highly-blueshifted signal can be detected to large heights. The
velocity cross-sections for
X-pixel 6 show large velocities in both the bright point and the
expanding loop. The primary component of the profile shows blue-shifted
velocities of up to 20~\kms\ in the expanding loop.

The LOS velocities of the secondary component from X-pixel 6 are seen
to increase away from the bright point between Y-pixels 32 and 23. This
behavior suggests that plasma acceleration is occurring within the jet,
and it was also seen in the 9-Feb jet \citep{young-blowout} where it
was cited as evidence for the nonlinear Alfv\'en waves predicted from
the jet model of \citet{pariat09}. Although not shown, we note that
comparable velocity increases are seen in X-pixels 4 and 5, and a
smaller, though still significant, velocity increase is seen from
X-pixel 7.

\begin{figure}
 \begin{center}
  \includegraphics[width=7cm]{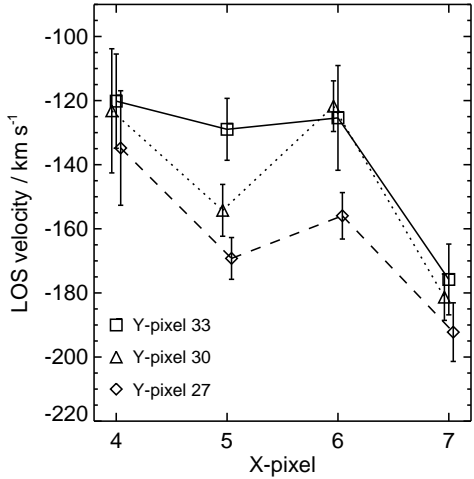} 
 \end{center}
\caption{LOS velocities of the secondary Gaussian component to
  \ion{Fe}{xii} \lam195.12 plotted across the width of the jet
  (X-position), for three different Y-positions. The three sets of
  points are plotted slightly offset from each other in order to aid
  viewing.}\label{fig.twist}
\end{figure}

As the jet is spatially resolved in the transverse direction to the
jet axis, then it is possible to consider the variation of LOS
velocity across the jet. Three-dimensional models of jets
\citep[e.g.,][]{pariat09} predict some degree of twisting motion which
would be reflected in the LOS velocities being larger on one side of
the jet than the other. Figure~\ref{fig.twist} shows velocity
cross-sections through the jet at three Y-positions (refer to
Figure~\ref{fig.eis}e for the locations of the pixels). Although
larger velocities are seen on the right-side of the jet (X-pixel 7)
compared to the left-side of the jet, consistent with twist, there is
not a monotonic variation of LOS velocity across the jet axis. This
may indicate more complex motions within the jet, but we also caution
that the data points are taken over a period of four minutes and there
is likely a significant time dependence that can not be accounted
for. Despite this it is clear that the sensitivity of the EIS velocity
measurements is sufficient to resolve twisting motions and there is
some evidence that the present jet is twisted with the right-side
showing a larger LOS speed compared to the left-side by 60~\kms. This
would correspond to anti-clockwise rotation if looking down the axis
of the jet towards the bright point.

If we interpret the velocity difference between X-pixels 4 and 7 as
representing the rotational motions around the axis of the jet, then
we can estimate the angular speed and period as $9.0\times
10^{-3}/\sin\phi$~rad~s$^{-1}$ and 690\,$\sin\phi$~s, respectively,
where $\phi$ is the angle between the jet axis and the LOS to the
observer. Given the location on the solar surface of the bright point
and the morphology of the images, we consider it likely that
$\phi\approx 45$--90$^\circ$, giving $\sin\phi\approx0.7$--1.0. The
cylindrical radius of the jet is taken to be 4.5\as = 3300~km based on
the width of the EIS velocity feature. \citet{shen11} found values of $11.1\times
10^{-3}$~rad~s$^{-1}$ and 564~s for an active region jet,
\citet{chen12} values of $14.1\times
10^{-3}$~rad~s$^{-1}$ and 452~s for a coronal hole jet, and
\citet{hong13} values of $13.3\times
10^{-3}$~rad~s$^{-1}$ and 475~s for another coronal hole jet, so the
present values are comparable. We note that these authors derived the
twist parameters through high-cadence imaging by tracking the
transverse motions of discrete structures within the jets.

\begin{figure}
 \begin{center}
  \includegraphics[width=8cm]{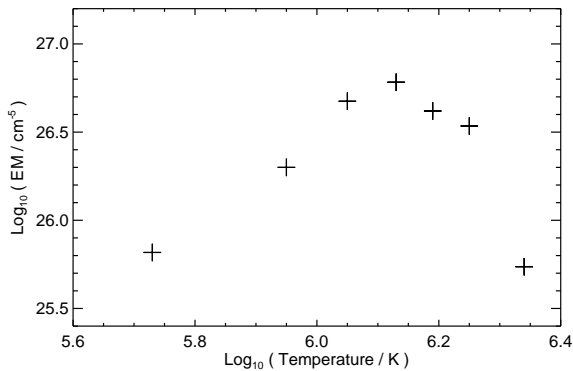} 
 \end{center}
\caption{Column emission measure values for the bright point, derived
  from EIS intensity measurements of lines of \ion{Fe}{viii--xiii} and
  \ion{Fe}{xv}.}\label{fig.em}
\end{figure}

The temperature reached by the bright point during the event was
investigated by 
creating an average spectrum from the EIS data. Thirty-five spatial
pixels in the brightest part of the \ion{Fe}{xii} \lam195.12
image were selected and averaged using the IDL routine
EIS\_MASK\_SPECTRUM \citep{eis_sw16}, and Gaussians were fit to the
emission lines in the spectrum using the IDL routine
SPEC\_GAUSS\_EIS. The line intensities are given in
Table~\ref{tbl.ints}. The spectrum showed a weak 
\ion{Fe}{xv} \lam284.16 line, but 
\ion{Fe}{xvi} \lam262.99 was not present. The densities derived from
the \ion{Fe}{xii} \lam186.88/\lam195.12 and \ion{Fe}{xiii}
\lam203.82/\lam202.04 ratios are $\log\,N_{\rm e}=8.90$ and 8.86,
respectively. We refer the reader to \citet{young09-dens} for details on applying
these diagnostics to EIS data, and we note that atomic data were taken
from version~7.1 of the CHIANTI atomic database
\citep{dere97,chianti71}. Emission lines of
\ion{Fe}{viii--xiii} and \ion{Fe}{xv} are available from the data-set,
and the intensities of the lines have been converted to column
emission measure values using the method described in Section~3.1 of
\citet{tripathi10}. The emission measure values are plotted in
Figure~\ref{fig.em}.
The coronal iron abundance value of
\citet{schmelz12} was used, and the density was assumed to be constant
with temperature with a value of $\log\,N_{\rm e}=8.9$. The emission
measure curve peaks at $\log\,T=6.13$ (\ion{Fe}{xi}), and falls
sharply between \ion{Fe}{xiii} and \ion{Fe}{xv}. This is consistent
with the lack of \ion{Fe}{xvi} \lam262.99 emission, and also the very
weak signal in the \axvi\ channel. We can thus place a constraint on
the maximum temperature reached in the bright point as $\log\,T\approx
6.3$ (2~MK). Previously \citet{doschek10} presented an analysis of a
coronal hole bright point that produced a jet and found similar plasma
parameters to those found here. No \ion{Fe}{xvi} emission was seen,
the differential emission measure curve peaked at $\log\,T=6.1$, and
the density (derived from three diagnostics) was $\log\,N_{\rm
  e}=8.85$--9.05.

For the jet, a $2\times 10$ block of pixels at the tip of the jet was
averaged to yield a single spectrum. Another $2\times 10$ block of
pixels to the right of the jet was also averaged to yield a background
spectrum that was subtracted from the jet spectrum, thus isolating the
jet emission from the background. 
This
demonstrated that the jet only emits in \ion{Fe}{xi--xiii}, with 
no significant emission from
\ion{Fe}{viii--x} or \ion{Fe}{xv}. The intensities for
the \ion{Fe}{xi--xiii} lines  are shown in
Table~\ref{tbl.ints}. The density implied from the \ion{Fe}{xii}
\lam186.88/\lam195.12 diagnostic is $\log\,N_{\rm
  e}=8.10^{+0.14}_{-0.17}$. Isothermal temperatures derived from
\ion{Fe}{xii} \lam195.12/\ion{Fe}{xi} \lam188.22 and \ion{Fe}{xiii}
\lam202.04/\ion{Fe}{xii} \lam195.12 are $\log\,T=6.21$ and 6.23,
respectively. (These values were derived using the CHIANTI atomic
database.)
The LOS velocities of \ion{Fe}{xi--xiii} -- measured relative to the
line positions in the background spectrum -- are $-150\pm 9$, $-162\pm
2$ and $-180\pm 6$~\kms\ for \ion{Fe}{xi} \lam192.81, \ion{Fe}{xii}
\lam195.12 and \ion{Fe}{xiii} \lam202.04, suggesting that there is a
temperature dependence of the outflow speed.

\begin{table}
  \caption{EIS line intensities.}\label{tbl.ints}
  \begin{center}
    \begin{tabular}{lccc}
      \hline
\noalign{\smallskip}
           &&\multicolumn{2}{c}{Intensity (\ecss)} \\
\cline{3-4}
\noalign{\smallskip}
Ion & Line & BP  & Jet \\
      \hline
\noalign{\smallskip}
\ion{Fe}{viii} &\lam185.21 & $32.1\pm 1.0$ & ---\\
\ion{Fe}{ix}   &\lam197.86 & $13.8\pm 0.3$  & ---\\
\ion{Fe}{x}    &\lam184.54 & $78.7\pm 1.6$ & --- \\
\ion{Fe}{xi}   &\lam188.22 & $137.5\pm 1.6$ & $10.5\pm 1.0$ \\
\ion{Fe}{xii}  &\lam195.12 & $195.0\pm 1.1$ & $24.9\pm 0.8$ \\
               &\lam186.88 & $47.4\pm 1.0$ & $2.1\pm 0.5$  \\
\ion{Fe}{xiii} &\lam202.04 & $84.3\pm 1.5$ & $24.8\pm 2.3$ \\
               &\lam203.82 & $62.3\pm 1.9$ & --- \\
\ion{Fe}{xv}   &\lam284.16 & $24.5\pm 1.4$ & --- \\
      \hline
    \end{tabular}
  \end{center}
\end{table}

The jet parameters can be compared with previous EIS observations of
coronal hole jets studied by \citet{doschek10} and
\citet{young-blowout}. \citet{doschek10} found temperatures of
$\log\,T=6.05$--6.12 using a \ion{Fe}{xi}/\ion{Fe}{xii} ratio,
measured at four heights along their jet. \citet{young-blowout} used a
\ion{Fe}{xi}/\ion{Fe}{xii} ratio to derive a temperature of
$\log\,T=6.12$. \citet{doschek10} measured densities of $\log\,N_{\rm
  e}=7.70$ and 8.85 at two locations in their jet, and
\citet{young-blowout} found a density of $\log\,N_{\rm e}=8.44$. These
results suggest that coronal hole jets typically release plasma close
to the formation temperature of \ion{Fe}{xi} and \ion{Fe}{xii}
($\log\,T=6.1$--6.2) and with a low density typical of coronal hole or
quiet Sun plasma.

\section{Summary}\label{sect.summary}

The key properties of the 8-Feb jet are as follows:

\begin{itemize}
\item The jet is an expanding, loop-shaped structure that emerges
  from the negative polarity end of a small magnetic bipole.
\item The jet is triggered by cancellation of a weak, positive
  polarity magnetic feature that moves towards the negative pole of
  the bipole.
\item The bright point does not get hotter than 2.0~MK during the
  event, and the coronal density is $7.6\times 10^8$~cm$^{-3}$.
\item The jet duration is 25~minutes, and the bright point is not
  significantly disrupted by the jet occurrence (the intensity returns
  to the pre-jet levels, and the magnetic field strength remains
  similar). 
\item The jet extends 30~Mm (projected distance) above the bright
  point and the temperature of the ejected plasma is
  1.7~MK; the density is in the range 0.9 to 1.7 $\times$
  $10^{8}$~cm$^{-3}$; and  absorption in the \aix\ channel suggests that chromospheric
  plasma is also ejected.
\item LOS speeds increase with distance from the bright point,
  reaching values up to  250~\kms.
\item Evidence from the spectroscopic LOS velocities is found for
  twisted motions in the jet, and an angular speed of $9-12 \times
  10^{-3}$~rad~s$^{-1}$ is derived.
\end{itemize}

Comparing with the 9-Feb jet analyzed by \citet{young-blowout}, the key
difference lies with the magnetic field evolution: for the 9-Feb jet
the two dominant polarities canceled with each other, whereas for the
8-Feb jet a parasitic polarity emerges near the negative polarity footpoint,
and cancels with it. This difference is reflected in the evolution of
the bright point during the event: the base arch of the 9-Feb jet is
blown open by the jet, and a number of very intense kernels are seen;
yet for the 8-Feb jet the base arch remains and no kernels were seen. 
Despite these differences, the temperature, density and
velocity of the
ejected plasma are similar between the two events.

On-disk observations of coronal holes afford an excellent opportunity
for comparing different types of coronal jet, as both the  magnetic
and coronal evolution can be studied. The results here and in
\citet{young-blowout} suggest that
blowout jets can behave quite differently depending on the particular
magnetic field geometry and evolution. Further categorization perhaps
based on the types of magnetic field interaction (majority--majority
or minority--majority cancelation) may be appropriate.

\bigskip

The authors acknowledge funding from National Science Foundation grant
AGS-1159353, and we thank the referee for valuable
comments. P.R.Y. thanks ISSI for financial support to attend the 2014
International Team Meeting ``Understanding Solar Jets and their Role
in Atmospheric Structure and Dynamics'' (PI: N.-E.~Raouafi), and he
thanks the participants for useful discussions.
\hinode\ is a Japanese mission developed and launched by 
ISAS/JAXA, with NAOJ as domestic partner and NASA and
STFC (UK) as international partners. It is operated by
these agencies in co-operation with ESA and NSC (Norway). SDO is a
mission for NASA's Living With a Star 
program, and data are provided courtesy of NASA/SDO and the AIA and
HMI science teams.

\bibliographystyle{apj}
\bibliography{myrefs}

\begin{thebibliography}{}

\bibitem[\protect\citeauthoryear{{Chen}, {Zhang}, \& {Ma}}{{Chen}
  et~al.}{2012}]{chen12}
{Chen}, H.-D., {Zhang}, J.,  \& {Ma}, S.-L. 2012, Research in Astronomy and
  Astrophysics, 12, 573

\bibitem[\protect\citeauthoryear{{Dere} et~al.}{{Dere} et~al.}{1997}]{dere97}
{Dere}, K.~P., {Landi}, E., {Mason}, H.~E., {Monsignori Fossi}, B.~C.,  \&
  {Young}, P.~R. 1997, \aaps, 125, 149

\bibitem[\protect\citeauthoryear{{Doschek} et~al.}{{Doschek}
  et~al.}{2010}]{doschek10}
{Doschek}, G.~A., {Landi}, E., {Warren}, H.~P.,  \& {Harra}, L.~K. 2010, \apj,
  710, 1806

\bibitem[\protect\citeauthoryear{{Hong} et~al.}{{Hong} et~al.}{2011}]{hong11}
{Hong}, J., {Jiang}, Y., {Zheng}, R., {Yang}, J., {Bi}, Y.,  \& {Yang}, B.
  2011, \apjl, 738, L20

\bibitem[\protect\citeauthoryear{{Hong} et~al.}{{Hong} et~al.}{2013}]{hong13}
{Hong}, J.-C., {Jiang}, Y.-C., {Yang}, J.-Y., {Zheng}, R.-S., {Bi}, Y., {Li},
  H.-D., {Yang}, B.,  \& {Yang}, D. 2013, Research in Astronomy and
  Astrophysics, 13, 253

\bibitem[\protect\citeauthoryear{{Landi} et~al.}{{Landi}
  et~al.}{2013}]{chianti71}
{Landi}, E., {Young}, P.~R., {Dere}, K.~P., {Del Zanna}, G.,  \& {Mason}, H.~E.
  2013, \apj, 763, 86

\bibitem[\protect\citeauthoryear{{Lee} et~al.}{{Lee} et~al.}{2013}]{kslee13}
{Lee}, K.-S., {Innes}, D.~E., {Moon}, Y.-J., {Shibata}, K., {Lee}, J.-Y.,  \&
  {Park}, Y.-D. 2013, \apj, 766, 1

\bibitem[\protect\citeauthoryear{{Liu} et~al.}{{Liu} et~al.}{2011}]{liu11}
{Liu}, C., {Deng}, N., {Liu}, R., {Ugarte-Urra}, I., {Wang}, S.,  \& {Wang}, H.
  2011, \apjl, 735, L18

\bibitem[\protect\citeauthoryear{{Moore} et~al.}{{Moore}
  et~al.}{2010}]{moore10}
{Moore}, R.~L., {Cirtain}, J.~W., {Sterling}, A.~C.,  \& {Falconer}, D.~A.
  2010, \apj, 720, 757

\bibitem[\protect\citeauthoryear{{Moore} et~al.}{{Moore}
  et~al.}{2013}]{moore13}
{Moore}, R.~L., {Sterling}, A.~C., {Falconer}, D.~A.,  \& {Robe}, D. 2013,
  \apj, 769, 134

\bibitem[\protect\citeauthoryear{{Nistic{\`o}} et~al.}{{Nistic{\`o}}
  et~al.}{2009}]{nistico09}
{Nistic{\`o}}, G., {Bothmer}, V., {Patsourakos}, S.,  \& {Zimbardo}, G. 2009,
  \solphys, 259, 87

\bibitem[\protect\citeauthoryear{{O'Dwyer} et~al.}{{O'Dwyer}
  et~al.}{2010}]{odwyer10}
{O'Dwyer}, B., {Del Zanna}, G., {Mason}, H.~E., {Weber}, M.~A.,  \& {Tripathi},
  D. 2010, \aap, 521, A21

\bibitem[\protect\citeauthoryear{{Pariat}, {Antiochos}, \& {DeVore}}{{Pariat}
  et~al.}{2009}]{pariat09}
{Pariat}, E., {Antiochos}, S.~K.,  \& {DeVore}, C.~R. 2009, \apj, 691, 61

\bibitem[\protect\citeauthoryear{{Schmelz} et~al.}{{Schmelz}
  et~al.}{2012}]{schmelz12}
{Schmelz}, J.~T., {Reames}, D.~V., {von Steiger}, R.,  \& {Basu}, S. 2012,
  \apj, 755, 33

\bibitem[\protect\citeauthoryear{{Schmieder} et~al.}{{Schmieder}
  et~al.}{2013}]{schmieder13}
{Schmieder}, B., et~al. 2013, \aap, 559, A1

\bibitem[\protect\citeauthoryear{{Shen} et~al.}{{Shen} et~al.}{2012}]{shen12}
{Shen}, Y., {Liu}, Y., {Su}, J.,  \& {Deng}, Y. 2012, \apj, 745, 164

\bibitem[\protect\citeauthoryear{{Shen} et~al.}{{Shen} et~al.}{2011}]{shen11}
{Shen}, Y., {Liu}, Y., {Su}, J.,  \& {Ibrahim}, A. 2011, \apjl, 735, L43

\bibitem[\protect\citeauthoryear{{Shibata} \& {Uchida}}{{Shibata} \&
  {Uchida}}{1986}]{shibata86}
{Shibata}, K.,  \& {Uchida}, Y. 1986, \solphys, 103, 299

\bibitem[\protect\citeauthoryear{{Shimojo} et~al.}{{Shimojo}
  et~al.}{1996}]{shimojo96}
{Shimojo}, M., {Hashimoto}, S., {Shibata}, K., {Hirayama}, T., {Hudson}, H.~S.,
   \& {Acton}, L.~W. 1996, \pasj, 48, 123

\bibitem[\protect\citeauthoryear{{Tripathi} et~al.}{{Tripathi}
  et~al.}{2010}]{tripathi10}
{Tripathi}, D., {Mason}, H.~E., {Del Zanna}, G.,  \& {Young}, P.~R. 2010, \aap,
  518, A42

\bibitem[\protect\citeauthoryear{{Young} \& {Muglach}}{{Young} \&
  {Muglach}}{2013}]{young-blowout}
{Young}, P.,  \& {Muglach}, K. 2013, ArXiv e-prints

\bibitem[\protect\citeauthoryear{{Young}}{{Young}}{2012}]{eis_sw16}
{Young}, P.~R. 2012, EIS Software Note No.~16, ver.~2.4

\bibitem[\protect\citeauthoryear{{Young} et~al.}{{Young}
  et~al.}{2009}]{young09-dens}
{Young}, P.~R., {Watanabe}, T., {Hara}, H.,  \& {Mariska}, J.~T. 2009, \aap,
  495, 587

\end{thebibliography}

\end{document}